\documentclass[aps,prb, 
superscriptaddress,amsmath,amssymb,showpacs,twocolumn]{revtex4-1}

\usepackage{graphicx}
\usepackage{hyperref}
\usepackage{epsfig}
\usepackage{color}
\usepackage{psfrag}
\usepackage{float}
\usepackage{bbold}
\usepackage{tikz}

\newcommand{\rem}[1]{}
\newcommand{\refe}[1]{~(\ref{#1})}

\newcommand{\enlev}{\varepsilon_0}

\begin{document}

\title{Build-up of Vibron-Mediated Electron Correlations in Molecular Junctions}
\author{R. Avriller}
\affiliation{Univ. Bordeaux, CNRS, LOMA, UMR 5798, F-33405 Talence, France}
\author{R. Seoane Souto}
\affiliation{Departamento de F\'{i}sica Te\'{o}rica de la Materia Condensada,\\
Condensed Matter Physics Center (IFIMAC) and Instituto Nicol\'{a}s Cabrera,
Universidad Aut\'{o}noma de Madrid E-28049 Madrid, Spain}
\author{A. Mart\'in-Rodero}
\affiliation{Departamento de F\'{i}sica Te\'{o}rica de la Materia Condensada,\\
Condensed Matter Physics Center (IFIMAC) and Instituto Nicol\'{a}s Cabrera,
Universidad Aut\'{o}noma de Madrid E-28049 Madrid, Spain}
\author{A. Levy Yeyati}
\affiliation{Departamento de F\'{i}sica Te\'{o}rica de la Materia Condensada,\\
Condensed Matter Physics Center (IFIMAC) and Instituto Nicol\'{a}s Cabrera,
Universidad Aut\'{o}noma de Madrid E-28049 Madrid, Spain}
\date{\today}

\begin{abstract}
We investigate on the same footing the time-dependent electronic transport properties and vibrational dynamics of a molecular junction. 
We show that fluctuations of both the molecular vibron displacement and the electronic current across the junction undergo damped oscillations towards the steady-state. 
We assign the former to the onset of electron tunneling events assisted by vibron-emission.
The time-dependent build-up of electron-hole correlations is revealed as a departure of the charge-transfer statistics from the generalized-binomial one after a critical time $t_c$.
The phonon-back action on the tunneling electrons is shown to amplify and accelerate this build-up mechanism.
\end{abstract}


\maketitle

%
\textit{Introduction}.--
The scaling of electronic junctions down to the molecule or single-atom size \cite{stipe1998coupling,xu2003measurement,van1996adjustable} is known to 
suffer from some limitations. 
%
To cite but a few of them: experiments are poorly reproducible implying statistical averaging on many samples \cite{tal2008electron}, transport characteristics are highly dependent on geometry and chemical nature of the tip or substrate \cite{lee1999single}, and mechanical properties of the junction are degraded by voltage-induced heating \cite{ioffe2008detection}, up to reaching mechanical instability and final break-down \cite{huang2007single}.
The previous limitations involve interaction between electronic and vibrational degrees of freedom of the molecular junction. 
It is thus of both fundamental and practical importance for molecular electronics to better understand the impact of electron-phonon (e-ph) excitations on electronic transport at the nanoscale. 
%

%
Typical signatures of e-ph interactions are measured in the conductance $G(V)$ 
characteristics as peaks or dips \cite{galperin2004inelastic}, 
appearing each time the bias-voltage $V$ crosses
the inelastic threshold $\hbar\omega_0/e$, with $\omega_0$ the local-vibron frequency, $\hbar$ the reduced Planck constant and $e$ the electron charge. 
The analysis of the position and width of these inelastic features \cite{galperin2004inelastic} contains information about the e-ph matrix elements, the excited vibron frequencies and lifetimes \cite{stipe1998coupling}. 
%
More recently, signatures of electron-vibron excitations were also reported on shot-noise $S(V)$ characteristics \cite{kumar2012detection}, 
revealing complementary information about electronic correlations mediated by vibron excitation.
%
%
This extensive experimental activity has been supported by great theoretical efforts, the aim of which has been to clarify the fundamental mechanism of electron-tunneling assisted by vibron-emission and its impact on quantum transport \cite{galperin2006resonant,mitra2004phonon,viljas2005electron,galperin2004inelastic,paulsson2005modeling,PhysRevB.73.075428,egger2008vibration,frederiksen2007inelastic,PhysRevB.82.165441,PhysRevB.86.155411}.
%
%
Despite all these efforts, the understanding of electron-electron, electron-vibron interactions and the role of electronic coherence at the nanoscale remains mainly limited to the stationary (time-independent) transport regime.
%

%
This topic has experienced a revival with the recent development of single-electron sources \cite{feve2007demand},
which allow controlled injection of well-defined single-electron excitations in atomic point contacts. 
This has opened new avenues for probing the short-time response of a nanojunction, in the range 1-10 ns \cite{feve2007demand}. 
%
Further improvements in designing broadband and low-noise detectors has been later reported, with the first measurement of thermal decay of current-fluctuations at ultrashort time scales 10-100 ps \cite{thibault2015pauli}. 
%
%
It is thus timely to develop new
theoretical tools 
bridging the gap between molecular electronics and ultrafast quantum electronics \cite{perfetto2018cheers,Tang_2017}.
Such approaches should enable the computation of the mean current \cite{perfetto2015transient,muhlbacher2008real,wang2011numerically}, current-current noise and higher-order cumulants of the current fluctuations \cite{esposito2009nonequilibrium,tang2014waiting,1367-2630-20-8-083039},
including the non-Markovian character of electronic tunneling at low-temperature.
For those reasons, the understanding of interaction effects on time-dependent transport is still a challenging issue. 
%

In this Rapid Communication, we develop a compact methodology based on nonequilibrium Green functions (NEGF) \cite{keldysh1965diagram,caroli1971direct,kamenev2011field} for probing on the same footing time-dependent electronic current-fluctuations and vibron dynamics of a molecular junction. 
%
We follow the junction dynamics 
from short time-scales given by the inverse electronic tunneling rate $1/\Gamma$, to a longer time-window characterized by the vibron-mode inverse  damping rate $1/\gamma_d$ and by electronic-current transient oscillations of period $2\pi\hbar/\left( eV \pm \hbar\omega_0 \right)$.  
%
We show the departure of the charge-transfer statistics from the non-interacting generalized-binomial distribution \cite{hassler2008wave}, at a critical time associated to the build-up of vibron-mediated electron correlations.
%

%
\textit{Microscopic model}.--Our approach is based on a microscopic Hamiltonian for the molecular junction 
$H(t) = H_{\mathcal{M}} + H_{\mathcal{R}} + H_{\mathcal{T}}(t)$ \cite{holstein1959studies,mitra2004phonon,galperin2006resonant}, with
\begin{eqnarray}
H_{\mathcal{M}} &=& \enlev n_d
+ \hbar\omega_0 a^\dagger a +
\lambda\left( a + a^\dagger \right) 
\left( n_d - \frac{1}{2} \right)
\label{Hamiltonian_2} \, , \\
H_{\mathcal{R}} &=& \sum_{r,k} \xi_{r,k} c^\dagger_{r,k} c_{r,k}
\label{Hamiltonian_3} \, ,\\
H_{\mathcal{T}}(t) &=& \sum_{r,k} \left\lbrace
t_{r,k}(t) c^\dagger_{r,k} d + t^*_{r,k}(t) d^\dagger c_{r,k}  
\right\rbrace
\label{Hamiltonian_4} \, .
\end{eqnarray}
Eq.\refe{Hamiltonian_2} describes a single electronic level of energy $\enlev$ and a local vibration mode of frequency $\omega_0$, with $d^\dagger$ ($a^\dagger$) the creation operator of an electronic (vibrational) excitation on the molecule. 
Electron-phonon interactions couple the position operator of the phonon mode (in units of its zero-point motion) $x = a + a^\dagger$ to the charge operator of the molecule $n_d \equiv d^\dagger d$, with coupling strength $\lambda$.
Eq.\refe{Hamiltonian_3} models the metallic left (L) and right (R) leads, with $c^\dagger_{r,k}$ the creation operator of an electronic excitation in the $r=L,R$ reservoir with energy $\xi_{r,k}$ and quasi-momentum $k$.
The leads are supposed to be in thermal equilibrium at temperature $T$, and their respective chemical potentials to be maintained under a symmetric voltage-drop $\mu_{L(R)}=\pm eV/2$.
Finally, Eq.\refe{Hamiltonian_4} describes the tunneling of electrons from  lead $r$ to the molecular level, with the rate $\Gamma_r\left( \omega \right)=\pi/\hbar \sum_k |t_{r,k}|^2\delta\left( 
\omega - \xi_{r,k}\right)$.
Within the wide-band approximation, the rates are evaluated at the Fermi energy $\Gamma_r\left( \omega \right)\approx \Gamma_r\left( E_F \right)\equiv \Gamma_r$, thus resulting in a
total tunneling rate $\Gamma=\Gamma_L+\Gamma_R$.
In order to probe the transient dynamics of charge-transfer across the junction, the tunneling hoppings $t_{r,k}(t)=t_{r,k}\theta(t)$ are switched-on at the initial time $t=0$, 
where $\theta(t)$ is the Heavyside step-function.
Typical experimental parameters for molecular junctions are \cite{smit2002measurement,tal2008electron,kumar2012detection}
:
$\hbar\Gamma \approx 1 \mbox{ eV}$, $\hbar\omega_0 \approx 10 \mbox{ meV}-100 \mbox{ meV}$,
$\left(\lambda/\hbar\Gamma\right)^2 \approx 1-5\%$ and $T \approx 4.2 \mbox{ K}$.
In the following, we adopt units such that $e=1$, $\hbar=1$ and the Boltzmann constant $k_B=1$.
The e-ph coupling $\left( \lambda/\Gamma \right)^2\approx 20\%$ and phonon frequency $\omega_0/\Gamma \approx 0.5$ are taken a bit larger than in usual experiments in order to achieve fast-enough relaxation.
%

%
%
%
%
%
\begin{figure}[tbh]
\includegraphics[width=1\linewidth]{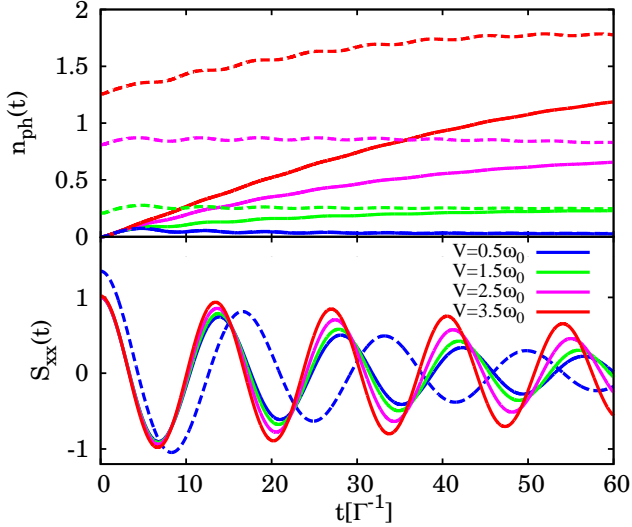}
\caption{%
Top panel: Time-dependent phonon
population $n_{ph}(t)$ for two different initial conditions $n_{ph}(0)=0$ (solid lines) and $n_{ph}(0) = n^{st}_{ph}$ (dashed lines). 
Lower panel: Vibron displacement fluctuations 
$S_{xx}(t)$ for the initial condition $n_{ph}(0)=0$ (plain curves), 
compared to the relaxation of a classical harmonic oscillator (dashed-blue line) in the case $V=0.5\omega_0$ and $n_{ph}(0) = n^{st}_{ph} \approx 0$.
Common to both panels: $\Gamma=1$, $\Gamma_L=\Gamma_R$, $\epsilon_0=0$, $\omega_0=0.5$, $\lambda=0.45$, $T=0$,  
$n_d(0)=0$ and $V=0.5$, $1.5$, $2.5$, $3.5\omega_0$.}
\label{fig:Fig1}
\end{figure}
\textit{NEGF approach}.--We are interested in the full-counting statistics (FCS) of electron tunneling \cite{levitov1996electron,bagrets2003full,levitov2004counting,belzig2005full}, which provides complete information about current-fluctuations.
The central quantity in a FCS analysis is the 
quasi-probability distribution $P_q(t)$ that $q$ charges are transferred across the molecular junction between the initial
and final measurement times $0$ and $t$. 
The related moment generating function (MGF) 
$\mathcal{Z}(\chi,t)=\sum_{q\in \mathbb{Z}} P_q(t) e^{iq \chi}$
and cumulant generating function (CGF) 
$\mathcal{F}(\chi,t)= \ln \mathcal{Z}(\chi,t)$,
generate upon n-successive derivations with respect to the counting-field $\chi$, the $n^{th}$ moment $M_n$  and cumulant $C_n$ of the distribution $P_q(t)$ respectively.
%
%
The CGF is expressed as \cite{esposito2009nonequilibrium,tang2014waiting}
\begin{eqnarray}
\frac{\partial}{\partial \chi} \mathcal{F}(\chi,t) &=&
- \mbox{tr} \left\lbrace \frac{\partial \mathbf{\Sigma}_{T,\chi}}{\partial \chi} \mathbf{G}_\chi \right\rbrace
\label{Gogolin_1} \, ,
\end{eqnarray}
which recovers the stationary limit \cite{gogolin2006towards,kamenev2011field}. 
Eq.\refe{Gogolin_1} involves the time-dependent tunneling self-energy $\Sigma_{T,\chi}(1,2)$, and the nonequilibrium Green functions (NEGFs) \cite{keldysh1965diagram,caroli1971direct,kamenev2011field}
of the molecular level
$G_\chi(1,2)=-i\left\langle T_K d(1)d^\dagger(2)\right\rangle_\chi$ and vibron mode $D_{\chi}(1,2)=-i\left\langle  T_K x(1)x(2)\right\rangle_\chi$.
We adopt the short-hand notations for the time $t_{1(2)}\equiv 1(2)$, 
and the time-ordering operator $T_K$, on the Keldysh contour $\mathcal{C}_K$.
We write in bold symbol any matrix in the discretized contour. 
%
Notice that the dimension of the bold matrices increases linearly with time $t$.
The NEGFs are evaluated with the counting-field $\chi(t)$ included into the hopping terms $t_{r,k}(t)\equiv t_{r,k}e^{i\chi_r(t)}$ \cite{levitov2004counting,gogolin2006towards}, with $\chi_{r}(t)=\pm s_r\chi/2$ for $t$ on the forward (backward) branch of $\mathcal{C}_K$ and $s_r=1(-1)$ for $r=L(R)$.
%

%
We evaluate Eq.\refe{Gogolin_1} 
within the Random Phase Approximation (RPA) \cite{urban2010nonlinear,novotny2011nonequilibrium,utsumi2013full},
for which the molecular level and vibron NEGFs fulfill the following equations
\begin{eqnarray}
\mathbf{G}_\chi &\approx& \mathbf{G}_{0\chi} + 
\mathbf{G}_{0\chi} \mathbf{\Sigma}_{eph,\chi} \mathbf{G}_{0\chi}\label{Gogolin_2} \, , \\
\mathbf{D}_\chi &=& \left\lbrace \mathbf{d}_0^{-1}
- \mathbf{\Pi}_{\chi}\right\rbrace^{-1} \label{Gogolin_3} \, ,
\end{eqnarray}
with $\mathbf{G}_{0\chi}=\left\lbrace \mathbf{g}^{-1}
- \mathbf{\Sigma}_{T,\chi}\right\rbrace^{-1}$ the NEGF of the molecular level coupled to the leads but not interacting with the vibron mode, $\mathbf{g}$ the NEGF of the isolated level, and $\mathbf{d}_0$ the bare vibron propagator.
%
%
The electron self-energy $\mathbf{\Sigma}_{eph,\chi}$ in Eq.\refe{Gogolin_2} is the sum of an Hartree (H) 
term $\mathbf{\Sigma}_{H,\chi}$, plus an exchange (XC) contribution $\mathbf{\Sigma}_{XC,\chi}$,
while $\mathbf{\Pi}_{\chi}$ in Eq.\refe{Gogolin_3} is the vibron self-energy, given by
\begin{eqnarray}
\Sigma_{H,\chi}(1,2) &=& \lambda^2 \delta_K(1,2) \int_{\mathcal{C}_K} dt_3 n_{d,\chi}(3) d_0(1,3) \label{Gogolin_4} \, , \\
\Sigma_{XC,\chi}(1,2) &=& i\lambda^2 G_{0\chi}(1,2) D_\chi(1,2) \label{Gogolin_5} \, , \\
\Pi_{\chi}(1,2) &=& -i\lambda^2 G_{0\chi}(1,2)G_{0\chi}(2,1)\label{Gogolin_6} \, ,
\end{eqnarray}
where $\delta_K(1,2)$ is the delta-function defined on the Keldysh contour, and $n_{d,\chi}(3)$ the counting-field dependent population of the molecular level.
Consistently with the RPA, the electronic NEGF is truncated at second-order in the e-ph coupling strength $\left(\lambda/\Gamma\right)^2$ \cite{avriller2009electron,schmidt2009charge,haupt2009phonon}, while the vibron  propagator is 
obtained after resummating a whole class of dominant ring-diagrams \cite{utsumi2013full}.
%
%
%
Eq.\refe{Gogolin_1} to \refe{Gogolin_6} are the basis of our approach.
We solve them numerically, after discretizing the Keldysh contour \cite{souto2015transient,1367-2630-20-8-083039}.
Within RPA, taking into account only $\mathbf{\Sigma}_{XC,\chi}$ in Eq.\refe{Gogolin_2} ($\mathbf{\Sigma}_{H,\chi}$ gives a smaller contribution associated to displacement currents), Eq.\refe{Gogolin_1} can be integrated exactly and provides the following expression for the MGF:
$\mathcal{Z}(\chi,t) \approx \mbox{det}\left\lbrace \tilde{\mathbf{G}}_{\chi=0}\tilde{\mathbf{G}}_\chi^{-1}\right\rbrace /
\sqrt{\mbox{det}\left\lbrace \mathbf{D}_{\chi=0}\mathbf{D}_\chi^{-1}\right\rbrace}$ \cite{utsumi2013full}.
%
We have checked numerically that within RPA and for our range of parameters, the continuity equation for the electronic current is fulfilled. 
%

%
%
\begin{figure}
\includegraphics[width=1\linewidth]{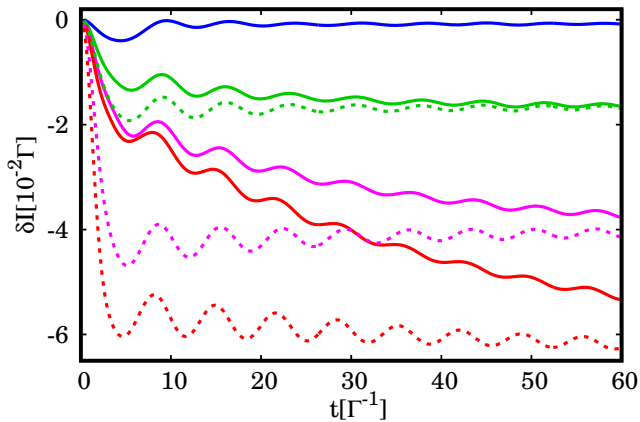}
\caption{%
Top panel : Time-dependence of the excess current $\delta I(t)=I(t)-I_0(t)$ with respect to the current $I_0(t)$ in the non-interacting case ($\lambda=0$), for two different initial conditions $n_{ph}(0)=0$
(solid lines) and $n_{ph}(0) = n^{st}_{ph}$
(dashed lines).
Parameters and legends are the same as in Fig.\ref{fig:Fig1}.
}
\label{fig:Fig2}
\end{figure}
%
%
\textit{Vibron dynamics}.--We focus first on the average phonon population $n_{ph}(t)\equiv \left\langle a^\dagger(t)a(t) \right\rangle$.
%
%
We show in Fig.\ref{fig:Fig1} (top panel) the time-evolution of $n_{ph}(t)$ for a symmetric junction $\Gamma_L=\Gamma_R$ 
with a resonant molecular level $\enlev = 0$, corresponding to a perfectly transmitting junction.
At the initial time, the molecular level is unoccupied, $n_d(0)=0$, while the vibron mode is in its ground state, i.e. $n_{ph}(0)=0$ (plain-curves).
Consistently with a rate equation description \cite{viljas2005electron,paulsson2005modeling}, we find that the vibron occupation slowly relaxes towards the steady-state value $n^{st}_{ph}=\left( V - \tilde{\omega}_0 \right)\theta\left( V - \tilde{\omega}_0 \right)/4\tilde{\omega}_0$, 
with a dissipation rate $\gamma_d=2\lambda^2\tilde{\omega}_0/\pi\Gamma^2$ and renormalized (softened) phonon frequency $\tilde{\omega}_0 = \omega_0 - 2\lambda^2/\pi\Gamma$.
We estimate $\tilde{\omega}_0 \approx 76\% \omega_0$ and
$\gamma_d \approx 5.4\%\Gamma$, implying a relaxation time 
$1/\gamma_d \approx 18/\Gamma$ which is consistent with the low-voltage numerical curves.
For higher voltages ($V\geq 2.5\omega_0$), inelastic electron-tunneling events heat up the phonon mode \cite{galperin2004inelastic,paulsson2005modeling}, while the dissipation rate becomes voltage-dependent. 
As expected, the relaxation is 
faster for the initial condition $n_{ph}(0) = n^{st}_{ph}$
closer to the steady-state (dashed-curves). 
%

The fluctuations of the vibron displacement 
$S_{xx}(t)\equiv\mbox{ Re}\left\langle x(t)x(0) \right\rangle$ are 
shown for $n_{ph}(0)=0$ in Fig.\ref{fig:Fig1} (lower panel).
In the case $V=0.5\omega_0$ (blue curve), $S_{xx}(t)$ exhibits damped-oscillations with period $2\pi/\tilde{\omega}_0$, and decoherence time $1/\gamma_{x}\approx 2/\gamma_d$, in good agreement with the relaxation of a classical harmonic oscillator (dashed-blue curve):
$S_{xx}(t) \approx
\left(1+2 n^{st}_{ph}\right)e^{-\gamma_x t}
\left\lbrace \cos\left(\tilde{\omega}_0 t\right)
+ \frac{\gamma_x}{\tilde{\omega}_0} \sin\left(\tilde{\omega}_0 t\right) \right\rbrace$\footnote{
This expression is derived by neglecting the energy-dependence of the phonon self-energy in Eq.\refe{Gogolin_6} \cite{viljas2005electron,novotny2011nonequilibrium}. This approximation at the pole is consistent with the rate-equation, and is valid when the broadening of the phonon spectrum $\gamma_x$ is very weak compared to the phonon frequency $\omega_0$.}.
We notice a phase-shift between the plain and dashed curves due to the retardation of the vibron in responding to the tunneling electrons.
The value of $\gamma_x$ is found larger than half the dissipation rate as a result of additional dephasing induced by elastic tunneling of electrons \cite{avriller2018bistability}.
%

%
\begin{figure}
\includegraphics[width=1\linewidth]{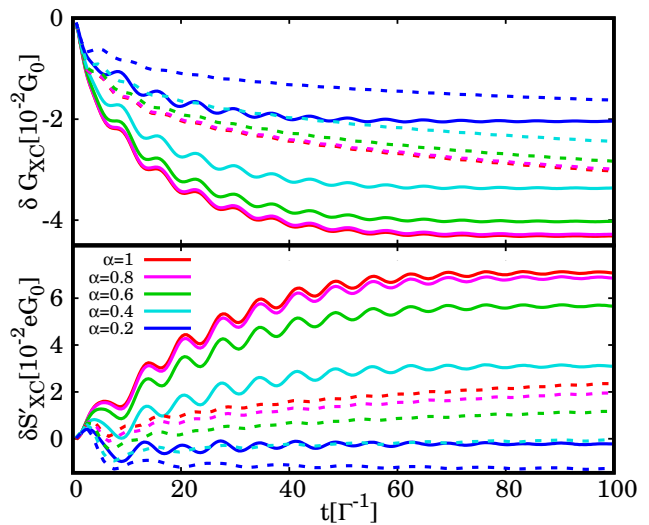}
\caption{
Top panel: Time-dependence of the excess
conductance $\delta G(t)$ (plain-curves) at the inelastic threshold $V \simeq \omega_0$, in units of the quantum of conductance $G_0=e^2/h$. 
Lower panel: Voltage-derivative of the excess current-fluctuations $\delta S'(t)$ at the inelastic threshold.
The dashed-curves are obtained by evaluating Eq.\refe{Gogolin_5} with  the bare phonon propagator $d_{0}(1,2)$.
Common to both panels: $\Gamma=1$, $\epsilon_0=0$, $\omega_0=0.5$, $V=\omega_0$, $\lambda=0.45$, $T=0$, $n_d(0)=0$, $n_{ph}(0)=0$ and
$\alpha\equiv\Gamma_L/\Gamma_R=1$, $0.8$, $0.6$, $0.4$, $0.2$.
}
\label{fig:Fig3}
\end{figure}
%
\textit{Electronic transport}.--We consider now the average symmetrized current across the junction $I(t)\equiv \frac{d}{dt} C_1(t)$, and time-derivative of the related symmetrized charge-fluctuations $S(t)= \frac{d}{dt} C_2(t)$.
We define the excess current $\delta I(t) = I(t) - I_0(t)$ and excess current-fluctuations $\delta S(t)= S(t) - S_0(t)$ with respect to the current $I_0(t)$ and current-fluctuations $S_0(t)$ in the non-interacting case ($\lambda=0$).
We show in Fig.\ref{fig:Fig2}, the time-evolution of $\delta I(t)$, for the same parameters as in Fig.\ref{fig:Fig1}.
%
We find that $\delta I(t)$ oscillates and relaxes toward the steady-state inelastic current:
$\delta I^{st} \approx - \left( \lambda/\Gamma \right)^2 \left\lbrace 2 n_{ph}^{st} V + \left( V - \tilde{\omega}_0 \right)
\theta  \left( V - \tilde{\omega}_0 \right)\right\rbrace/2\pi$ \cite{paulsson2005modeling,egger2008vibration,haupt2009phonon}.    
The transient oscillations
with period $\approx 2\pi/\left( V \pm \tilde{\omega}_0\right)$, 
are associated to the maintained phase-coherence during vibron-assisted inelastic tunneling events.
When approaching the steady-state, the gradual loss of coherence results in a
power-law decay of the oscillation amplitude.
We also probe the dependence with the junction transmission $\tau=4\alpha/\left(1+\alpha\right)^2$, by changing the ratio between the tunneling rates $\alpha=\Gamma_L/\Gamma_R$.
We show in Fig.\ref{fig:Fig3} (top-panel) the excess conductance $\delta G(t)= \frac{\partial}{\partial V} \delta I(t)$ evaluated at $V\approx\omega_0$ (plain curves).
As predicted by bare second-order perturbation theory \cite{kim2014inelastic}, $\delta G(t)$ is negative for arbitrary values of $\alpha$ (with fixed $\enlev=0$).
The difference between plain (RPA) and 
dashed (bare second-order) curves measures the impact of the vibron-heating mechanism.
We find that the onset of a non-equilibrium vibron population in the junction tends to lower the stationary conductance while amplifying the transient oscillations of $\delta G(t)$.
A similar conclusion is drawn in Fig.\ref{fig:Fig3} (lower-panel) for
the voltage-derivative of the excess current-noise $\delta S'(t)= \frac{\partial}{\partial V} \delta S(t)$ at $V\approx\omega_0$.
We remark an over-amplification of $\delta S'(t)$ at $\tau=1$, due to phonon back-action \cite{urban2010nonlinear,novotny2011nonequilibrium,utsumi2013full}.
A quench of the transient oscillations and a change of sign of $\delta S'(t)$ is observed at $\tau=1/2$ ($\alpha\approx 0.17$), as
the dominant scattering channel changes from inelastic tunneling of electrons to elastic tunneling with emission-reabsorption of a vibron  \cite{kim2014inelastic}. 
We have checked that Fig.\ref{fig:Fig3} is qualitatively unchanged for the initial condition $n_d(0)=1$, except for small differences at very short times $t\le 10/\Gamma$ where the transient dynamics is slowed-down by the suppressed charge-fluctuations of the occupied dot. 
%

%
%
\begin{figure}
\includegraphics[width=1\linewidth]{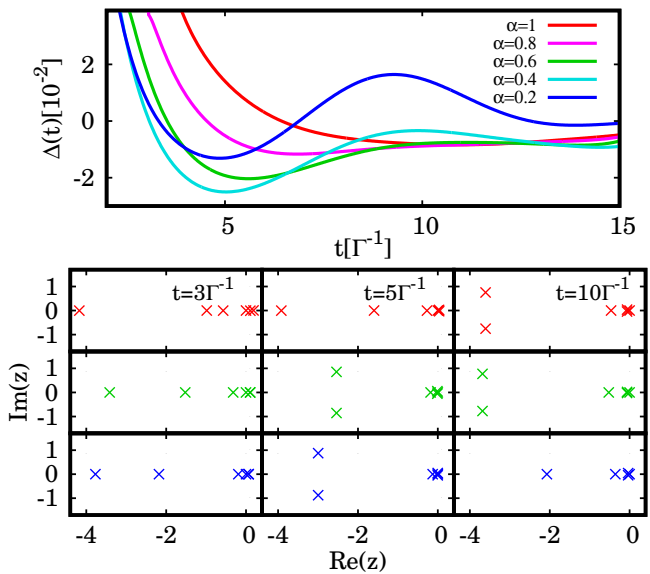}
\caption{
Top panel: Discriminant $\Delta(t)=P^2_0(t)-4P_1(t)P_{-1}(t)$ of the short-time approximation to the 
MGF $\mathcal{Z}(z,t) = P_{1}(t)z+P_0(0)+P_{-1}(t)z^{-1}$,
with $z=e^{i\chi}$.
The lower panels show the 
position of the zeros of the MGF for $\alpha=1$ (upper red row), $\alpha=0.6$ (middle green row) and $\alpha=0.2$ (lower blue row), with increasing times from left to right columns.
Other parameters are the same as in Fig.\ref{fig:Fig3}.
}
\label{fig:Fig4}
\end{figure}
%
%
\textit{Zeros of the MGF}.--In order to characterize charge fluctuations beyond the first two cumulants, we investigate the analytical properties of the MGF $\mathcal{Z}(z,t)=\sum_{q\in\mathbb{Z}}P_q(t) z^q\equiv\prod_{j} \left( z-z_j \right)/\left( 1-z_j \right)$ as a function of $z \equiv e^{i\chi}$, extended to the full complex plane.
The zeros $z_j$ of the MGF are either real or come in complex-conjugate pairs.
For a non-interacting fermionic system, the MGF factorizes to $\mathcal{Z}(z,t)\equiv \prod_{j} \left\lbrace 1 + p_j \left( z - 1 \right)\right\rbrace$ \cite{PhysRevB.75.205329,abanov2008allowed} with $p_j \in \left\lbrack0,1\right\rbrack$ being the probability of the binomial tunneling process, so that the zeros 
$z_j\equiv 1 - 1/p_j$ lie on the negative real axis.
Any departure of the zeros from the real axis is thus a direct signature of electron correlations \cite{PhysRevB.92.155413,utsumi2013full,PhysRevB.96.165444}.
%
%
%
Similar studies were reported in the context 
of dynamical phase transitions, for the real-time evolution of bulk systems \cite{PhysRevLett.110.135704} or in relation to full-counting statistics \cite{PhysRevLett.110.050601}, for which the zeros of the MGF were later determined experimentally \cite{PhysRevLett.118.180601}.
At short-times ($t<15/\Gamma$), the MGF is dominated by single-electron tunneling events, i.e. $\mathcal{Z}(z,t) \approx P_{1}(t)z+P_0(0)+P_{-1}(t)z^{-1}$,
where $P_{1}(t)$ and $P_{-1}(t)$ are the respective probabilities of forward and backward tunneling.
The sign of the discriminant $\Delta(t)=P^2_0(t)-4P_1(t)P_{-1}(t)$ controls the location of the zeros of $\mathcal{Z}(z,t)$ with respect to the real axis.
We present in Fig.\ref{fig:Fig4} the computed zeros of the full MGF as a function of time (lower-panel) and the corresponding behavior of the 
discriminant $\Delta(t)$ (upper-panel), for the same parameters as in Fig.\ref{fig:Fig3}.
At short times $(t \lessapprox 1/\Gamma)$, 
the zeros lie on the negative real axis, for arbitrary $\alpha$, as expected for non-interacting systems \cite{PhysRevB.75.205329,abanov2008allowed}. 
After some time $(t > 1/\Gamma)$, the electrons have tunneled on the molecule and emitted a vibron.
The onset of e-ph interactions results into a merging of the zeros of the MGF at a critical time $t_c$, and their later splitting off the real axis for $t>t_c$.
The time $t_c$ coincides with the change of sign of the discriminant $\Delta(t)$ from positive to negative, thus
proving that the splitting of the zeros is due to a departure from the generalized binomial distribution of non-interacting electrons \cite{hassler2008wave}.
We interpret this behavior as arising from correlations between single-electron inelastic tunneling events and inelastic back-scattering ones (single-hole transmission).
For our available time-window and range of parameters, 
the phonon back-action mechanism leads to an amplification of the 
electron-hole correlations, and thus to a shorter $t_c$ compared to the case of bare second-order perturbation theory.
At half transmission $\tau\approx 0.5$ ($\alpha\approx 0.2$), 
the zeros first split, then merge again at time $t \approx 6.2/\Gamma$, and finally stay on the negative real axis.
This quench of electron-hole correlations happens as the dominant scattering process changes from vibron-mediated inelastic to elastic tunneling of electrons, 
thus resulting in a FCS closer to the one of a non-interacting junction.
%

\textit{Conclusion}.--In this Rapid Communication, we have investigated on the same footing the time-dependent transport properties and vibrational dynamics of a molecular junction.
We have shown that the fluctuations of the vibron displacement exhibit damped oscillations toward the steady state similar to the relaxation of a classical harmonic oscillator.
The short-time dynamics of current and current-fluctuations
exhibit voltage-dependent oscillations, due to both the mean-field reorganization of molecular charges and to the onset of inelastic scattering.
This short-time dynamics is mainly due the building-up of vibron-mediated electron-hole correlations, 
the signature of which is revealed as a splitting of the zeros of the MGF off the real axis, at a critical time $t_c$.
The phonon back-action mechanism tends to amplify the electron-hole correlations, as well as the transient oscillations of electronic current-fluctuations.
We believe that our work provides a first step to investigate the onset of many-body correlations in electronic transport, including the possibility to analyze vibron-mediated dynamical phase transitions \cite{utsumi2013full}, when reaching the stationary regime.
%
Recent progress in the THz spectroscopy of photo-currents in molecular junctions \cite{du2018terahertz} and of photon-assisted shot-noise in graphene \cite{PhysRevLett.116.227401}, constitute an alternative and promising route to investigate the subtle interplay between electrons and vibron dynamics at ultrashort time scales $\sim 1-10 \mbox{ ps}$, along the lines proposed in this paper. 
\\

R.A. acknowledges support from R\'egion de la Nouvelle Aquitaine,  the Transnational Common Laboratory "QuantumChemPhys: Theoretical
Chemistry and Physics at the Quantum Scale", and the Agence Nationale de la Recherche, project CERCa, ANR-18-CE30-0006.
R.S.S., A.L.Y. and A.M.R. acknowledge financial support by Spanish MINECO (Grants No. FIS2014-55486-P and FIS2017-84860-R), and the Mar\'{\i}a de Maeztu Program (Grant No. MDM-2014-0377). 
%

\bibliography{Molecular_Junction}

\end{document}